\documentclass[a4paper,11pt]{article}
\usepackage{jheppub}
\usepackage[T1]{fontenc} 
\usepackage{graphicx}
\usepackage{amsmath}
\usepackage[caption=false]{subfig}

\usepackage{breqn}

\def\cL{  {\cal L}  }
\newcommand{\bea}{\begin{eqnarray}}
\newcommand{\eea}{\end{eqnarray}}
\newcommand{\be}{\begin{equation}}
\newcommand{\ee}{\end{equation}}
\newcommand{\dr}{{\rm d}}
\newcommand{\dd}{{\rm d}}
\newcommand{\nn}{\nonumber}

\newcommand{\pa}{\partial}
\newcommand{\ep}{\epsilon}

\newcommand{\zb}{\bar{z}}

\allowdisplaybreaks[4]

\preprint{HU-EP-16/20, HU-MATH 2016-13}
\title{Evaluating four-loop conformal Feynman integrals by $D$-dimensional
differential equations}

\author[a]{Burkhard Eden,}
\author[b]{Vladimir A.\ Smirnov}
\affiliation[a]{
Institut f\"{u}r Mathematik und Physik, Humboldt-Universit{\"a}t zu Berlin, Zum gro{\ss}en Windkanal 6, 12489 Berlin, Germany}
\affiliation[b]{Skobeltsyn Institute of Nuclear Physics of Moscow State 
University, 119992 Moscow, Russia}
\emailAdd {eden@math.hu-berlin.de}
\emailAdd {smirnov@theory.sinp.msu.ru}

\abstract{
We evaluate a four-loop conformal integral, i.e. an integral over four four-dimensional
coordinates, by turning to its dimensionally regularized version and applying
differential equations for the set of the corresponding 213 master integrals.
To solve these linear differential equations we follow the strategy suggested by Henn
and switch to a uniformly transcendental basis of master integrals.
We find a solution to these equations up to weight eight in terms of multiple polylogarithms.
Further, we present an analytical result for the given four-loop conformal integral
considered in four-dimensional space-time in terms of single-valued harmonic polylogarithms.
As a by-product, we obtain analytical results for all the other 212 
master integrals within 
dimensional regularization, i.e. considered in $D$ dimensions.
}


\begin{document}

\maketitle
\flushbottom

\section{Introduction}
 
Correlation functions of half-BPS operators 
in ${\cal N}=4$ super Yang-Mills (SYM) theory --- in particular of the stress-tensor multiplet ---
have been extensively studied because their strong coupling regime is accessible via the AdS/CFT correspondence. Subsequently, the integrability of the model's spectrum problem was discovered, and in an initially unrelated effort, many results were obtained for scattering amplitudes and a dual set of Wilson loops. These two developments were brought together by the construction of an integrable system \cite{Alday:2009dv,Basso:2013vsa} for the so-called remainder function in amplitudes.

The stress-tensor correlators came back to center stage when it became clear that they can act as generating objects for both scattering amplitudes and the dual polygonal Wilson loops \cite{Alday:2010zy,Eden:2010zz}. It is then a natural question whether these correlators can also 
be analyzed from an integrable systems perspective. In \cite{Basso:2015zoa} such ideas have been put forward for general three-point functions in weak coupling perturbation theory. 
Unfortunately, comparison to perturbative ``data''  
is possible only indirectly because it is a fairly hard task to obtain exact field theory results for non-trivial three-point functions at higher loops. However, three-point couplings for two half-BPS and one twist operator are available from OPE limits of half-BPS four-point functions.
The explicit result for the two-loop four-point stress-tensor correlator from a decade ago \cite{Eden:2000mv,Bianchi:2000hn} 
has been a guideline for the construction of the ``hexagon proposal'' of \cite{Basso:2015zoa}, and more recent 
work \cite{Eden:2011we,Eden:2012rr,Drummond:2013nda,Chicherin:2015edu} for the three-loop part of the four-point 
function has been successfully compared to the hexagon prediction \cite{Eden:2015ija,Basso:2015eqa}; a vital test of the proposal.

At the next order it is not yet clear how to handle the hexagon due to problems with a double pole. On the field 
theory side, the integrand of the four-point function of stress-tensor multiplets has been 
elaborated in \cite{Eden:2011we,Eden:2012tu,Ambrosio:2013pba,Bourjaily:2015bpz} up to eight loops. 
It takes the form of a kinematic factor \cite{Eden:2000bk} times a sum of scalar conformal integrals 
in a propagator representation. At four loops (and beyond), the largest part of the integrals has not yet 
been evaluated: there are 26 genuine four-loop integrals in the planar part of the correlator 
(so the part relevant to integrability), of which five can be related to the ladder with four 
rungs by flip identities on subintegrals. One further integral could be solved in \cite{Drummond:2013nda} 
as it obeys a Laplace equation, see below. The recent paper \cite{Goncalves:2016vir} considers the leading terms in an asymptotic expansion for the entire set of integrals.

A possible way to exactly evaluate conformal integrals is to follow the idea advocated in \cite{Brown:2008um}
and to try to perform integration over Feynman parameters in an appropriate order.
If it turns out that there is an order in which the dependence of the denominator of integrand
on the Feynman parameters is linear then the whole integral can be solved in terms of
multiple polylogarithms. This strategy was successfully applied for example in 
\cite{Panzer:2013cha,Panzer:2014gra,vonManteuffel:2014qoa,Brown:2012ia,Schnetz:2013hqa}
and implemented as the computer code {\tt HyperInt} in \cite{Panzer:2014caa}. 

\begin{figure}[htb]
\begin{center}
\includegraphics[width=0.35\textwidth]{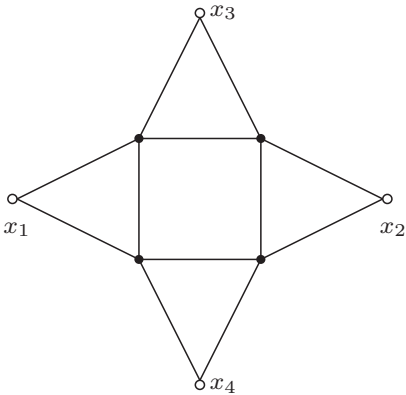}
\caption{A four-loop conformal integral.}
\label{fig1}
\end{center}
\end{figure}

We initiate here the study of the remaining 20 integrals by the method of differential equations, choosing at will 
the simplest looking diagram. The goal of this paper is thus to evaluate the coordinate-space Feynman integral 
associated with the graph of Fig.~\ref{fig1}.  
Although this integral is linearly reducible, in the sense of  \cite{Brown:2008um},
and an analytical result can be obtained with {\tt HyperInt} in \cite{Panzer:2014caa},
we are going to evaluate it with differential equations, keeping in mind that many cases in our set of 20 remaining four-loop conformal integrals will be linearly irreducible, although knowledge on polynomial reduction is constantly increasing. Indeed, it is helpful to be able to test our result against a different method.
 
The previous paper \cite{Drummond:2013nda}
chiefly aimed at three loops, but it also contains the aforementioned application of the Laplace equation to one of the 26 four-loop integrals.
Upon flipping a subintegral, one external vertex is connected to the rest of this diagram only by a single line. Acting by the operator $\Box=\frac{\partial}{\partial x_{\mu}}\frac{\partial}{\partial x^{\mu}}$,
where $x$ is the four-coordinate of this vertex,
is described in graph-theoretical language as the contraction of the corresponding line,
so that we easily obtain a differential equation in four dimensions.

\begin{figure}[htb]
\begin{center}
\includegraphics[width=0.35\textwidth]{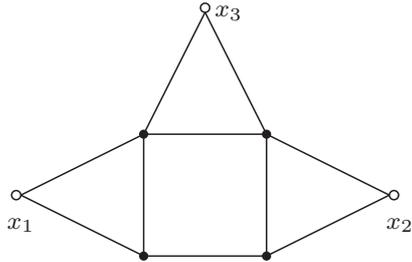}
\caption{A vertex integral obtained from our four-loop conformal integral.}
\label{fig2}
\end{center}
\end{figure}

In Figure 1, conformal invariance allows us to send the point $x_4$ to infinity to reduce to the Feynman integral whose graph is shown in Fig.~\ref{fig2}. 
Here two lines are incident to each of the three external vertices so that we cannot
derive a differential equation of the same type as in the previous case. 

The approach based on differential equations and integration by parts substantially enlarges any problem, because instead of a single integral one considers a family of master integrals closed under IBP relations.
A four-dimensional version of the method \cite{Caron-Huot:2014lda} 
is more economical with respect to the size of the basis of integrals. 
However, an obvious complication for our set of conformal integrals is that there are potentially ultraviolet and infrared divergences, and it is not straightforward 
to choose a subspace of Feynman integrals which includes a given conformal integral and other finite integrals (both in the ultraviolet and infrared sense) closed under four-dimensional integration by parts relations.

We believe that the strategy outlined in  \cite{Caron-Huot:2014lda} can be adjusted to the case
of conformal integrals as well. In this paper, however, we decided to employ the well-known straightforward
technique within dimensional regularization based on $D$-dimensional integration by parts \cite{Chetyrkin:1981qh} and differential equations \cite{Kotikov:1990kg,Kotikov:1991pm,Remiddi:1997ny,Gehrmann:1999as,Gehrmann:2000zt,Gehrmann:2001ck,Henn:2013pwa}.

On the one hand, we certainly make the situation more complicated, because dimensional regularization 
means involving a fairly large number of integrals. Indeed, as we will see shortly,
there are 213 master integrals in the corresponding extended family of Feynman integrals;
the integral of Fig.~\ref{fig2} is only one of them.
On the other hand, we obtain the possibility to apply the very well-known powerful machinery of differential equations.

To solve these linear differential equations we follow the strategy suggested by Henn \cite{Henn:2013pwa}
(see also \cite{Henn:2014qga}) and first applied in \cite{Henn:2013tua,Henn:2013woa,Henn:2013nsa}
and then in many other papers. To this end we switch to a uniformly transcendental basis of master integrals.
We obtain a solution to these equations up to weight eight in terms of multiple polylogarithms \cite{Goncharov:1998kja}. An analytical result for the given four-loop conformal integral
considered in four-dimensional space-time is given in terms of harmonic polylogarithms \cite{Remiddi:1999ew} of weight eight. 
This result can naturally be represented in terms of single-valued harmonic polylogarithms \cite{BrownSVHPLs}.
As a by-product we obtain analytical results for all the other 212 master integrals considered in $D$
dimensions.
   
In the next section, we present definitions, describe master integrals and differential equations and
explain how we arrived at a canonical basis, in the sense of \cite{Henn:2013pwa}.
In Section~3, we solve the differential equations in our canonical basis and describe
our results. We discuss perspectives in the conclusions. 

\section{Master integrals and differential equations}

Since after going to $D=4-2\ep$ dimensions we have to deal with a complete set of integrals closed under 
integration by parts relations, we introduce eight more numerators (in the form of propagators) in addition to the ten existing propagators. Thus, we arrive at the following family of integrals
\bea
F_{a_1,\ldots,a_{18}} &=&
\int\ldots\int \frac{\dr^Dx_5 \, \dr^Dx_6 \, \dr^Dx_7\, \dr^Dx_8}
{[-x_5^2]^{a_1} [-x_6^2]^{a_2} [-(x_1 - x_5)^2]^{a_3}[-(x_1 - x_7)^2]^{a_4}[-(x_2 - x_6)^2]^{a_5}}
\nonumber \\ && \hspace*{-15mm}
\times \frac{
[-(x_2 - x_5)^2]^{-a_{11}}[-(x_1 - x_6)^2]^{-a_{12}}[-(x_2 - x_7)^2]^{-a_{13}}
[-(x_6 - x_7)^2]^{-a_{14}}[-x_7^2]^{-a_{15}}}
{ [-(x_2 - x_8)^2]^{a_6} [-(x_5 - x_6)^2]^{a_7} 
 [-(x_5 - x_7)^2]^{a_8}}
\nonumber \\ && \hspace*{-15mm}
\times \frac{[-(x_1 - x_8)^2]^{-a_{16}}[-(x_5 - x_8)^2]^{-a_{17}}[-x_8^2]^{-a_{18}}}{[-(x_6 - x_8)^2]^{a_9} [-(x_7 - x_8)^2]^{a_{10}}} \; .
\label{integrals}
\eea
The powers of the propagators (indices) $a_i$ are integer and $a_i\leq 0$ for $i\geq 11$.
Although the situation is of Euclidean type we prefer to deal with propagators in Minkowski space, with
$x^2=x_0^2-\vec{x}^2$ because some computer codes are oriented at Minkowski propagators.

This is a family of vertex Feynman integrals depending on $x_1,x_2$ and $x_3$. In the last formula translation 
invariance was used to set $x_3$ to zero. We put $(x_1-x_2)^2=-1$ and introduce standard conformal variables by
\bea
x_1^2&=&-z\bar{z}\;,  
x_2^2=-(1-z)(1-\bar{z})\;.
\label{confvar}
\eea

Using {\tt FIRE}  \cite{Smirnov:2008iw,Smirnov:2013dia,Smirnov:2014hma} combined with {\tt LiteRed} \cite{Lee:2012cn} 
we reveal 213 master integrals. In particular, there are four master integrals in the top sector, i.e. with all the first ten indices positive. The integral number 210 is the integral  $F_{1,\ldots,1,0,\ldots}$ equal to the conformal integral in 
Figure~2 which was our starting point. The full list of primary master integrals is present in a file attached to this submission.

 
The derivation of differential equations for a family of master integrals is a straightforward
procedure. We take derivatives of the master integrals in $z$ and $\bar{z}$ with the help of 
the package {\tt LiteRed}  \cite{Lee:2012cn} and then apply {\tt FIRE} to reduce the
resulting integrals to master integrals. As a result we obtain two systems of
linear differential equations.
\bea
\frac{\pa}{\pa z} f &=& A_1(z,\zb,\ep) f\;,
\\
\frac{\pa}{\pa \zb} f &=& A_2(z,\zb,\ep) f\;,
\eea
where $f$ is the vector of primary master integrals and $A_1,A_2$ are $213\times 213$-matrices.
 
We follow the strategy suggested in \cite{Henn:2013pwa} and turn to a new basis where the differential equations take the form
\bea
\frac{\pa}{\pa z} f &=&\ep  \bar{A}_1(z,\zb) f\;,
\label{canonicalde1}
\\
\frac{\pa}{\pa \zb} f &=& \ep \bar{A}_2(z,\zb) f\;,
\label{canonicalde2}
\eea
where the matrices $\bar{A}_1,\bar{A}_2$ are independent of $\ep$.
  
In differential form, we have
\bea \label{DEdifferentialform}
d\,f(z,\zb) = \epsilon\, ( d \, \tilde{A}(z,\zb)) \, f(z,\zb) \,.
\nn
\eea 
where
\begin{align}
\tilde{A}  = \sum_{k}   \tilde{A}_{\alpha_k} \, \log(\alpha_{k} ) \,.
\end{align}
The matrices $ \tilde{A}_{\alpha_k}$ are {\it constant} matrices and
the arguments of the logarithms $\alpha_{i}$ ({\it letters}) are  
functions of $z,\zb$. One calls this form of differential equations canonical.

In our case, the list of letters is
\be
\{z, 1 - z, \zb, 1 - \zb, -z + \zb, 1 - z - \zb, 1 - z \zb, 
 z + \zb - z \zb\}\;.
\label{alphabet}
\ee

In the case of one variable there exists an algorithm \cite{Lee:2014ioa} which provides
the possibility of arriving at a canonical basis\footnote{In fact, some parts of it can be applied in the
case of several variables as well.}. 
In our case there are two variables. We followed recipes formulated in \cite{Henn:2013pwa,Henn:2014qga} and successfully applied in \cite{Henn:2013tua,Henn:2013woa,Henn:2013nsa,Henn:2016men} and many other papers. In particular, one tries to choose basis integrals that have constant leading singularities\footnote{Leading singularities are multidimensional residues of the integrand. They determine rational factors in front of the otherwise logarithmic functions in explicit expressions for Feynman integrals.} \cite{Cachazo:2008vp}.
Sometimes we also used small additional rotations of
the basis to 'integrate out' terms with an $\ep^0$ dependence, as it was done in many cases, see e.g. 
\cite{Henn:2013tua,Henn:2013woa,Henn:2013nsa,Henn:2016men}.

We have implemented these recipes in a code which we successfully applied in the case at hand. 
A description of this code will be given in a future publication together with
results on more complicated four-loop conformal integrals. Our canonical basis is presented in an ancillary which is attached to this submission.

\section{Solving canonical differential equations}

We solve the linear system (\ref{canonicalde1},\ref{canonicalde2}), in a power expansion in $\ep$ according to the strategy described in detail in  \cite{Henn:2014lfa,Caola:2014lpa}. We solve the first linear system, Eq.~(\ref{canonicalde1}), 
which results into multiple polylogarithms of the argument $z$, up to unknown functions of $\zb$.

The multiple polylogarithms are defined recursively by 
\be\label{eq:Mult_PolyLog_def}
 G(a_1,\ldots,a_n;z)=\,\int_0^z\,\frac{\dd t}{t-a_1}\,G(a_2,\ldots,a_n;t)
\ee
with  $a_i, z\in \mathbb{C}$ and $G(z)=1$. In the special case where  $a_i=0$ for all i 
one has by definition
\be
G(0,\ldots,0;x) = \frac{1}{n!}\,\ln^n x \;.
\ee

Then we substitute this solution into the second system, Eq.~(\ref{canonicalde2}), check that the dependence on $z$ drops 
out and solve the resulting linear system depending only on $\zb$, up to constants of weights $w\leq 8$. 
To fix these $213\times 9$ unknown constants, we match our results for the canonical basis in terms of 
multiple polylogarithms to the leading order asymptotic behaviour of the solution of Eqs.~(\ref{canonicalde1},\ref{canonicalde2}) in the Euclidean limit $x_1 \to 0$, or equivalently $z,\zb\to 0$, found by solving the differential equations in the limit.

The corresponding terms of the expansion can be written in the well-known graph-theoretical language -- see, e.g., \cite{Smirnov:2002pj}. 
Alternatively, they can be described in the language of expansion by regions \cite{Beneke:1997zp,Smirnov:2002pj}. The crucial point in the matching is that the leading-order contributions are classified according to the power dependence on the small parameter of the limit, i.e. $\sim z \zb$. This parameter enters
with powers of the form $-k\ep$ where $k=0,1,2,3$, and this is seen both from the point of
view of differential equations and expansion by regions.

Upon fixing all the constants in our solution we obtain analytic results for all the 213 elements
of the canonical basis. These are presented in two files (for contributions up to
weight 7 and of weight 8, respectively) which can be downloaded from: \\  \url{http://theory.sinp.msu.ru/~smirnov/ci4}


Element number 210 of the basis is
\bea
f_{210}=\ep^8 (z-\zb)^2 F_{1,\ldots,1,0,\ldots}\;.
\eea
Its $\ep$-expansion starts from order $\ep^8$. We obtain the following result for the original conformal
integral
\bea
&& F_{1,\ldots,1,0,\ldots}= \frac{4}{(z - \bar z)^2} \times \label{our_result_210} \\
\bigl[ && - \cL_{\{3, 5\}} + \cL_{\{5, 3\}} - \cL_{\{1, 2, 5\}} + 
 \cL_{\{1, 3, 4\}} + \cL_{\{1, 4, 3\}} - \cL_{\{1, 5, 2\}}  \nonumber \\
 && - \cL_{\{2, 1, 5\}} + \cL_{\{2, 5, 0\}} + \cL_{\{2, 5, 1\}} - 
 \cL_{\{3, 1, 4\}} - \cL_{\{3, 4, 1\}} + \cL_{\{4, 1, 3\}} \nonumber \\
 && - \cL_{\{4, 3, 0\}} - \cL_{\{4, 3, 1\}} + \cL_{\{5, 1, 2\}} + 
 \cL_{\{5, 2, 1\}} - \cL_{\{1, 1, 1, 5\}} + \cL_{\{1, 1, 2, 4\}} \nonumber \\
 && - \cL_{\{1, 1, 4, 2\}} + \cL_{\{1, 1, 5, 0\}} + 
 \cL_{\{1, 1, 5, 1\}} - \cL_{\{1, 2, 3, 2\}} - 
 \cL_{\{1, 2, 4, 1\}} + 2 \cL_{\{1, 3, 1, 3\}} \nonumber \\
 && + \cL_{\{1, 4, 2, 1\}} - \cL_{\{1, 5, 0, 0\}} - 
 \cL_{\{1, 5, 1, 0\}} - \cL_{\{1, 5, 1, 1\}} - 
 \cL_{\{2, 1, 1, 4\}} - \cL_{\{2, 1, 2, 3\}} \nonumber \\
 && + \cL_{\{2, 1, 4, 0\}} + \cL_{\{2, 3, 2, 1\}} + 
 \cL_{\{2, 4, 1, 0\}} + \cL_{\{2, 4, 1, 1\}} - 
 \cL_{\{3, 1, 3, 0\}} - 2 \cL_{\{3, 1, 3, 1\}} \nonumber \\
 && + \cL_{\{3, 2, 1, 2\}} + \cL_{\{3, 3, 0, 0\}} + 
 \cL_{\{3, 3, 1, 0\}} + \cL_{\{4, 1, 1, 2\}} - 
 \cL_{\{4, 1, 2, 0\}} - \cL_{\{4, 2, 1, 0\}} \nonumber \\
 && - \cL_{\{4, 2, 1, 1\}} + \cL_{\{5, 1, 1, 1\}} - 
 \cL_{\{1, 1, 1, 2, 3\}} - \cL_{\{1, 1, 1, 3, 2\}} + 
 \cL_{\{1, 1, 1, 4, 0\}} + \cL_{\{1, 1, 2, 1, 3\}} \nonumber \\
 && + \cL_{\{1, 1, 2, 3, 0\}} + \cL_{\{1, 1, 2, 3, 1\}} - 
 \cL_{\{1, 1, 3, 1, 2\}} + \cL_{\{1, 1, 3, 2, 1\}} - 
 \cL_{\{1, 1, 4, 0, 0\}} + \cL_{\{1, 2, 1, 1, 3\}} \nonumber \\
 && - \cL_{\{1, 2, 1, 3, 1\}} - \cL_{\{1, 2, 3, 1, 1\}} + 
 \cL_{\{1, 3, 1, 1, 2\}} + \cL_{\{1, 3, 1, 2, 1\}} - 
 \cL_{\{1, 3, 2, 1, 0\}} - \cL_{\{1, 3, 2, 1, 1\}} \nonumber \\
 && - \cL_{\{1, 4, 1, 0, 0\}} - \cL_{\{1, 4, 1, 1, 0\}} - 
 \cL_{\{2, 1, 1, 1, 3\}} - \cL_{\{2, 1, 1, 3, 1\}} + 
 \cL_{\{2, 1, 3, 0, 0\}} + 2 \cL_{\{2, 1, 3, 1, 0\}} \nonumber \\
 && + \cL_{\{2, 1, 3, 1, 1\}} - \cL_{\{2, 2, 1, 2, 0\}} - 
 \cL_{\{2, 3, 0, 0, 0\}} - \cL_{\{2, 3, 1, 0, 0\}} + 
 \cL_{\{2, 3, 1, 1, 1\}} + \cL_{\{3, 1, 1, 1, 2\}} \nonumber \\
 && - \cL_{\{3, 1, 1, 2, 0\}} - \cL_{\{3, 1, 1, 2, 1\}} + 
 \cL_{\{3, 1, 2, 0, 0\}} - \cL_{\{3, 1, 2, 1, 1\}} + 
 \cL_{\{3, 2, 1, 0, 0\}} + \cL_{\{3, 2, 1, 1, 0\}} \nonumber \\
 && + \cL_{\{3, 2, 1, 1, 1\}} - \cL_{\{4, 1, 1, 1, 0\}} + 
 \cL_{\{1, 1, 1, 1, 3, 0\}} - \cL_{\{1, 1, 1, 2, 1, 2\}} + 
 \cL_{\{1, 1, 1, 3, 1, 0\}} + \cL_{\{1, 1, 2, 1, 2, 1\}} \nonumber \\
 && - \cL_{\{1, 1, 3, 0, 0, 0\}} - 2 \cL_{\{1, 1, 3, 1, 0, 0\}} - 
 \cL_{\{1, 1, 3, 1, 1, 0\}} + \cL_{\{1, 2, 1, 1, 1, 2\}} + 
 \cL_{\{1, 2, 1, 2, 0, 0\}} - \cL_{\{1, 2, 1, 2, 1, 1\}} \nonumber \\
 && + \cL_{\{1, 3, 0, 0, 0, 0\}} + \cL_{\{1, 3, 1, 0, 0, 0\}} - 
 \cL_{\{1, 3, 1, 1, 1, 0\}} - \cL_{\{2, 1, 1, 1, 2, 0\}} - 
 \cL_{\{2, 1, 1, 1, 2, 1\}} + \cL_{\{2, 1, 1, 2, 0, 0\}} \nonumber \\
 && + \cL_{\{2, 1, 1, 2, 1, 0\}} - \cL_{\{2, 1, 2, 0, 0, 0\}} + 
 \cL_{\{2, 1, 2, 1, 1, 0\}} + \cL_{\{2, 1, 2, 1, 1, 1\}} - 
 \cL_{\{2, 2, 1, 0, 0, 0\}} - \cL_{\{2, 2, 1, 1, 0, 0\}} \nonumber \\
 && - \cL_{\{2, 2, 1, 1, 1, 0\}} + \cL_{\{3, 1, 1, 1, 0, 0\}} + 
 \cL_{\{1, 1, 1, 1, 2, 0, 0\}} + \cL_{\{1, 1, 1, 1, 2, 1, 0\}} - 
 \cL_{\{1, 1, 1, 2, 0, 0, 0\}}  \nonumber \\ 
 && - \cL_{\{1, 1, 1, 2, 1, 0, 0\}} + \cL_{\{1, 1, 2, 0, 0, 0, 0\}} - \cL_{\{1, 1, 2, 1, 1, 0, 0\}} - 
 \cL_{\{1, 1, 2, 1, 1, 1, 0\}} + \cL_{\{1, 2, 1, 0, 0, 0, 0\}}  \nonumber \\
  && + \cL_{\{1, 2, 1, 1, 0, 0, 0\}}+ \cL_{\{1, 2, 1, 1, 1, 0, 0\}} - \cL_{\{2, 1, 1, 1, 0, 0, 0\}}
  + \cL_{\{1, 1, 1, 1, 0, 0, 0, 0\}} \ \bigr] \nonumber
\eea
which is, of course, of weight 8. In the last formula $\cL$ denotes a single-valued harmonic polylogarithm
\begin{equation}
\cL_{\{a_1,\ldots,a_8\}} = (-1)^{\sum a_i} G(a_1, \ldots a_8; z) + \sum c_{ij} \, G(\underline a_i;z) \, G(\underline a_j;\bar z)
\end{equation}
where $\underline a_i \cup \underline a_j$ has length 8 and $\underline a_j$ is never the empty word. The coefficients $c_{ij}$ are polynomials of multiple zeta values such that all branch cuts cancel. The entries in the weight vectors are in the set $\{0,1\}$ and we are using the ``condensed notation" $\ldots 0,0,0,1 \ldots = \ldots 4 \ldots$ etc. 
These are the SVHPL's introduced by Brown in \cite{BrownSVHPLs}.

Note that (\ref{our_result_210}) takes a much simpler form after flipping points $x_2 \leftrightarrow x_3$ (i.e. $z \to 1/z, \, \bar z \to 1/\bar z$) followed by $x_1 \leftrightarrow x_2$ (which implies $z \to 1-z, \, \bar z \to 1 - \bar z$): the pure function in the square bracket transforms as
\begin{equation}
[ \ldots ] \to -\cL_{\{3, 5\}} + \cL_{\{5, 3\}} + \cL_{\{2, 5, 0\}} - 
 \cL_{\{4, 3, 0\}} - \cL_{\{1, 5, 0, 0\}} + \cL_{\{3, 3, 0, 0\}} - 
 \cL_{\{2, 3, 0, 0, 0\}} + \cL_{\{1, 3, 0, 0, 0, 0\}}
\end{equation}

We checked our result (\ref{our_result_210}) by a numerical calculation with {\tt FIESTA} \cite{Smirnov:2015mct}, 
as well as those for some other elements in the basis.
Our result (\ref{our_result_210}) is in agreement with a calculation by different means \cite{Schnetz}
about which we knew in advance and with a calculation\footnote{Thanks to Erik Panzer for the comparison.}
based on {\tt HyperInt} \cite{Panzer:2014caa}.

\section{Conclusions}
 
To evaluate a four-loop conformal integral we applied powerful techniques designed for
dimensionally regularized Feynman integrals, although we still believe that one can develop an efficient purely four-dimensional
technique to evaluate conformal integrals.
We have also evaluated a family of four-loop vertex master integrals with all the 
differences of the external coordinates
off the light-cone. The corresponding dual momentum space Feynman integrals
are vertex integrals with all the end-points off the light-cone.
We believe that this is a first example of a complete calculation of such
a family of vertex integrals at the level of four loops.
We hope to report on further results on four-loop conformal integrals obtained with the help
of $D$-dimensional differential equations.

\vspace{0.2 cm}
{\em Acknowledgments.}
We thank Johannes Henn for various pieces of advice,
Oliver Schnetz for comparison of our results with his which were obtained by an independent technique, 
and Johannes Broedel and Claude Duhr for help in performing computer manipulations with
multiple polylogarithms.
We thank Paul Heslop, Vladimir Mitev, Erik Panzer and Oliver Schnetz for careful reading of the
draft of the paper.
VS is grateful to Matthias Staudacher for kind hospitality at the Humboldt University of Berlin
where this project started.
The work of VS was partially supported by the Alexander von Humboldt Foundation (Humboldt Forschungspreis). 
BE acknowledges support by SFB 647 of the DFG, and the Cluster of Excellence ``Image, Knowledge, Gestaltung'' at Humboldt-University Berlin, funded by the Excellence Initiative and DFG.

\bibliographystyle{JHEP}


\begin{thebibliography}{99}

\bibitem{Alday:2009dv}
  L.~F.~Alday, D.~Gaiotto and J.~Maldacena,
  JHEP {\bf 1109} (2011) 032
  doi:10.1007/JHEP09(2011)032
  [arXiv:0911.4708 [hep-th]].

\bibitem{Basso:2013vsa}
  B.~Basso, A.~Sever and P.~Vieira,
  Phys.\ Rev.\ Lett.\  {\bf 111} (2013) no.9,  091602
  doi:10.1103/PhysRevLett.111.091602
  [arXiv:1303.1396 [hep-th]].

\bibitem{Alday:2010zy}
  L.~F.~Alday, B.~Eden, G.~P.~Korchemsky, J.~Maldacena and E.~Sokatchev,
  JHEP {\bf 1109} (2011) 123
  doi:10.1007/JHEP09(2011)123
  [arXiv:1007.3243 [hep-th]].

\bibitem{Eden:2010zz}
  B.~Eden, G.~P.~Korchemsky and E.~Sokatchev,
  JHEP {\bf 1112} (2011) 002
  doi:10.1007/JHEP12(2011)002
  [arXiv:1007.3246 [hep-th]].

\bibitem{Basso:2015zoa}
  B.~Basso, S.~Komatsu and P.~Vieira,
  arXiv:1505.06745 [hep-th].

\bibitem{Eden:2000mv}
  B.~Eden, C.~Schubert and E.~Sokatchev,
  Phys.\ Lett.\ B {\bf 482} (2000) 309
  doi:10.1016/S0370-2693(00)00515-3
  [hep-th/0003096].

\bibitem{Bianchi:2000hn}
  M.~Bianchi, S.~Kovacs, G.~Rossi and Y.~S.~Stanev,
  Nucl.\ Phys.\ B {\bf 584} (2000) 216
  doi:10.1016/S0550-3213(00)00312-6
  [hep-th/0003203].

\bibitem{Eden:2011we}
  B.~Eden, P.~Heslop, G.~P.~Korchemsky and E.~Sokatchev,
  Nucl.\ Phys.\ B {\bf 862} (2012) 193
  doi:10.1016/j.nuclphysb.2012.04.007
  [arXiv:1108.3557 [hep-th]].

\bibitem{Eden:2012rr}
  B.~Eden,
  arXiv:1207.3112 [hep-th].

\bibitem{Drummond:2013nda}
  J.~Drummond, C.~Duhr, B.~Eden, P.~Heslop, J.~Pennington and V.~A.~Smirnov,
  JHEP {\bf 1308} (2013) 133
  doi:10.1007/JHEP08(2013)133
  [arXiv:1303.6909 [hep-th]].

\bibitem{Chicherin:2015edu}
  D.~Chicherin, J.~Drummond, P.~Heslop and E.~Sokatchev,
  arXiv:1512.02926 [hep-th].

\bibitem{Eden:2015ija}
  B.~Eden and A.~Sfondrini,
  JHEP {\bf 1602} (2016) 165
  doi:10.1007/JHEP02(2016)165
  [arXiv:1510.01242 [hep-th]].

\bibitem{Basso:2015eqa}
  B.~Basso, V.~Goncalves, S.~Komatsu and P.~Vieira,
  Nucl.\ Phys.\ B {\bf 907} (2016) 695
  doi:10.1016/j.nuclphysb.2016.04.020
  [arXiv:1510.01683 [hep-th]].

\bibitem{Eden:2012tu}
  B.~Eden, P.~Heslop, G.~P.~Korchemsky and E.~Sokatchev,
  Nucl.\ Phys.\ B {\bf 862} (2012) 450
  doi:10.1016/j.nuclphysb.2012.04.013
  [arXiv:1201.5329 [hep-th]].

\bibitem{Ambrosio:2013pba}
  R.~G.~Ambrosio, B.~Eden, T.~Goddard, P.~Heslop and C.~Taylor,
  JHEP {\bf 1501} (2015) 116
  doi:10.1007/JHEP01(2015)116
  [arXiv:1312.1163 [hep-th]].

\bibitem{Bourjaily:2015bpz}
  J.~L.~Bourjaily, P.~Heslop and V.~V.~Tran,
  Phys.\ Rev.\ Lett.\  {\bf 116} (2016) no.19,  191602
  doi:10.1103/PhysRevLett.116.191602
  [arXiv:1512.07912 [hep-th]].

\bibitem{Eden:2000bk}
  B.~Eden, A.~C.~Petkou, C.~Schubert and E.~Sokatchev,
  Nucl.\ Phys.\ B {\bf 607} (2001) 191
  doi:10.1016/S0550-3213(01)00151-1
  [hep-th/0009106].

\bibitem{Goncalves:2016vir}
  V.~Goncalves,
  arXiv:1607.02195 [hep-th].

\bibitem{Brown:2008um}
  F.~Brown,
  Commun.\ Math.\ Phys.\  {\bf 287} (2009) 925
  doi:10.1007/s00220-009-0740-5
  [arXiv:0804.1660 [math.AG]].
  
\bibitem{Panzer:2013cha}
  E.~Panzer,
  Nucl.\ Phys.\ B {\bf 874} (2013) 567
  doi:10.1016/j.nuclphysb.2013.05.025
  [arXiv:1305.2161 [hep-th]].
 
\bibitem{Panzer:2014gra}
  E.~Panzer,
  JHEP {\bf 1403} (2014) 071
  doi:10.1007/JHEP03(2014)071
  [arXiv:1401.4361 [hep-th]].

\bibitem{vonManteuffel:2014qoa}
  A.~von Manteuffel, E.~Panzer and R.~M.~Schabinger,
  JHEP {\bf 1502} (2015) 120
  doi:10.1007/JHEP02(2015)120
  [arXiv:1411.7392 [hep-ph]].

\bibitem{Brown:2012ia}
  F.~Brown and O.~Schnetz,
  arXiv:1208.1890 [math.NT].
  
\bibitem{Schnetz:2013hqa}
  O.~Schnetz,
  Commun.\ Num.\ Theor.\ Phys.\  {\bf 08} (2014) 589
  doi:10.4310/CNTP.2014.v8.n4.a1
  [arXiv:1302.6445 [math.NT]].
  
  \bibitem{Panzer:2014caa}
  E.~Panzer,
  Comput.\ Phys.\ Commun.\  {\bf 188} (2015) 148
  doi:10.1016/j.cpc.2014.10.019
  [arXiv:1403.3385 [hep-th]].
     
\bibitem{Caron-Huot:2014lda} 
  S.~Caron-Huot and J.~M.~Henn,
  JHEP {\bf 1406}, 114 (2014)
  [arXiv:1404.2922 [hep-th]].

\bibitem{Chetyrkin:1981qh}
K.~G.~Chetyrkin and F.~V.~Tkachov,
  Nucl.\ Phys.\ B {\bf 192} (1981) 159.
  
\bibitem{Kotikov:1990kg}
 A.~V.~Kotikov,
  Phys.\ Lett.\ B {\bf 254} (1991) 158.
  
\bibitem{Kotikov:1991pm}
  A.~V.~Kotikov,
  Phys.\ Lett.\ B {\bf 267} (1991) 123.

\bibitem{Remiddi:1997ny}
 E.~Remiddi,
  Nuovo Cim.\ A {\bf 110} (1997) 1435
  [hep-th/9711188].

\bibitem{Gehrmann:1999as}
T.~Gehrmann and E.~Remiddi,
  Nucl.\ Phys.\ B {\bf 580} (2000) 485
  [hep-ph/9912329].

\bibitem{Gehrmann:2000zt}
 T.~Gehrmann and E.~Remiddi,
  Nucl.\ Phys.\ B {\bf 601} (2001) 248
  [hep-ph/0008287].

\bibitem{Gehrmann:2001ck}
T.~Gehrmann and E.~Remiddi,
  Nucl.\ Phys.\ B {\bf 601} (2001) 287
  [hep-ph/0101124].

\bibitem{Henn:2013pwa}
  J.~M.~Henn,
  Phys.\ Rev.\ Lett.\  {\bf 110} (2013) 25,  251601
  [arXiv:1304.1806 [hep-th]].

\bibitem{Henn:2014qga}
  J.~M.~Henn,
  J.\ Phys.\ A {\bf 48} (2015) 15,  153001
  [arXiv:1412.2296 [hep-ph]].
    
\bibitem{Henn:2013tua}
  J.~M.~Henn, A.~V.~Smirnov and V.~A.~Smirnov,
  JHEP {\bf 1307} (2013) 128
  [arXiv:1306.2799 [hep-th]].
  
\bibitem{Henn:2013woa}
  J.~M.~Henn and V.~A.~Smirnov,
  JHEP {\bf 1311} (2013) 041
  [arXiv:1307.4083].
  
\bibitem{Henn:2013nsa}
  J.~M.~Henn, A.~V.~Smirnov and V.~A.~Smirnov,
  JHEP {\bf 1403} (2014) 088
  [arXiv:1312.2588 [hep-th]].
    
    
\bibitem{Goncharov:1998kja}
 A.~B.~Goncharov,
  Math.\ Res.\ Lett.\  {\bf 5} (1998) 497
  [arXiv:1105.2076 [math.AG]].
    
\bibitem{Remiddi:1999ew} 
  E.~Remiddi and J.~A.~M.~Vermaseren,
  Int.\ J.\ Mod.\ Phys.\ A {\bf 15}, 725 (2000)
  [hep-ph/9905237].
     
\bibitem{BrownSVHPLs}
F.~C.~S.~Brown, 
C.~R.~Acad.~Sci. Paris, Ser. {\bf I} (2004) 338.    
     
     
     
 
  
     
\bibitem{Smirnov:2008iw}
  A.~V.~Smirnov,
  JHEP {\bf 0810} (2008) 107
  [arXiv:0807.3243 [hep-ph]].
  
\bibitem{Smirnov:2013dia}
  A.~V.~Smirnov and V.~A.~Smirnov,
  Comput.\ Phys.\ Commun.\  {\bf 184} (2013) 2820
  [arXiv:1302.5885 [hep-ph]].
  
\bibitem{Smirnov:2014hma}
  A.~V.~Smirnov,
  Comput.\ Phys.\ Commun.\  {\bf 189} (2014) 182
  [arXiv:1408.2372 [hep-ph]].
  
\bibitem{Lee:2012cn}
R.~Lee, {\it {Presenting LiteRed: a tool for the Loop InTEgrals REDuction}},
  \href{http://xxx.lanl.gov/abs/1212.2685}{{\tt arXiv:1212.2685}}. 
 
\bibitem{Lee:2014ioa}
  R.~N.~Lee,
  JHEP {\bf 1504} (2015) 108
  [arXiv:1411.0911 [hep-ph]].
  
\bibitem{Henn:2016men}
  J.~M.~Henn, A.~V.~Smirnov, V.~A.~Smirnov and M.~Steinhauser,
  JHEP {\bf 1605} (2016) 066
  doi:10.1007/JHEP05(2016)066
  [arXiv:1604.03126 [hep-ph]]. 
  
\bibitem{Cachazo:2008vp} 
  F.~Cachazo,
  arXiv:0803.1988 [hep-th].

\bibitem{Henn:2014lfa}
  J.~M.~Henn, K.~Melnikov and V.~A.~Smirnov,
  JHEP {\bf 1405} (2014) 090
  [arXiv:1402.7078 [hep-ph]].
  
\bibitem{Caola:2014lpa}
  F.~Caola, J.~M.~Henn, K.~Melnikov and V.~A.~Smirnov,
  JHEP {\bf 1409} (2014) 043
  [arXiv:1404.5590 [hep-ph]].

  
\bibitem{Smirnov:2002pj}
V.~A. Smirnov, {\it {Applied asymptotic expansions in momenta and masses}},
  {\em Springer Tracts Mod. Phys.} {\bf 177} (2002) 1--262.


\bibitem{Beneke:1997zp}
M.~Beneke and V.~A. Smirnov, 
Nucl. Phys. {\bf B522} (1998) 321--344,
[hep-ph/9711391].

  
\bibitem{Smirnov:2015mct}
  A.~V.~Smirnov,
  Comput.\ Phys.\ Commun.\  {\bf 204} (2016) 189
  doi:10.1016/j.cpc.2016.03.013
  [arXiv:1511.03614 [hep-ph]].


\bibitem{Schnetz}
O, Schnetz, to be published.

\end{thebibliography}

\end{document}